\documentclass[sigconf]{acmart}

\usepackage[english]{babel}
\usepackage{blindtext}
\usepackage{url}
\usepackage{multirow}
\usepackage{balance}
\usepackage[ruled,linesnumbered]{algorithm2e}

\renewcommand\footnotetextcopyrightpermission[1]{} 
\setcopyright{none}

\acmDOI{}

\acmISBN{}

\acmPrice{}

\begin{document}
\begin{sloppypar}


\title{
PreConfig: A Pretrained Model for Automating Network Configuration
}



\author{Fuliang Li, Haozhi Lang, Jiajie Zhang, Jiaxing Shen, Xingwei Wang}



\begin{abstract}
Manual network configuration automation (NCA) tools face significant challenges in versatility and flexibility due to their reliance on extensive domain expertise and manual design, limiting their adaptability to diverse scenarios and complex application needs. This paper introduces PreConfig, an innovative NCA tool that leverages a pretrained language model for automating network configuration tasks. PreConfig is designed to address the complexity and variety of NCA tasks by framing them as text-to-text transformation problems, thus unifying the tasks of configuration generation, translation, and analysis under a single, versatile model. Our approach overcomes existing tools' limitations by utilizing advances in natural language processing to automatically comprehend and generate network configurations without extensive manual re-engineering. We confront the challenges of integrating domain-specific knowledge into pretrained models and the scarcity of supervision data in the network configuration field. Our solution involves constructing a specialized corpus and further pretraining on network configuration data, coupled with a novel data mining technique for generating task supervision data. The proposed model demonstrates robustness in configuration generation, translation, and analysis, outperforming conventional tools in handling complex networking environments. The experimental results validate the effectiveness of PreConfig, establishing a new direction for automating network configuration tasks with pretrained language models.

\end{abstract}

\maketitle

\section{Introduction}
Network configuration automation refers to using software tools and scripts to automatically configure, manage, and deploy network devices like routers, switches, and firewalls.
Leading technology companies such as Microsoft, Meta, and Alibaba have leveraged NCA tools to reduce manual efforts, improve efficiency, and decrease costs in network management~\cite{hoyan,aura,netcov,campion,jinjing}. 
By automating configuration tasks including configuration synthesis, translation, and analysis, these firms have achieved operational efficiencies, rapid infrastructure scaling, and consistent network controls as they expand globally.

Current NCA tools rely heavily on manual design that require extensive domain expertise and manual effort to codify network policies, objectives, and behaviors.
Configuration synthesis tools such as NetComplete~\cite{netcomplete} and Propane~\cite{propane} generate target configurations by solving manually defined objective functions. 
Configuration translation tools like Juniper2Cisco~\cite{junipertrans} and Huawei2Cisco~\cite{huaweitrans} transform vendor-specific configs using manually created templates. 
Lastly, configuration analysis tools including Batfish~\cite{batfish}, Config2Spec~\cite{config2spec} and Minesweeper~\cite{minesweeper} analyze configuration semantics based on expert-crafted text parsers and rules. 


Manual design-based NCA tools have limitations in versatility and flexibility. 
First, most of the tools are single-purpose, only automating individual tasks like synthesis or translation. 
Real-world NCA often requires combining multiple tools to perform different tasks. 
Using multiple disparate tools inevitably escalates learning and maintenance costs.
Second, it is challenging to extend these tools to support complex application scenarios due to inherent network and protocol complexity. 
For example, configuration synthesis and verification tools face scalability issues like state explosion in large networks~\cite{netcomplete}. 
\textit{Therefore, we raise the question can we develop a NCA solution that can address diverse scenarios and tasks?}

This paper provides an affirmative answer by proposing PreConfig, a pretrained model-based NCA tool that demonstrates versatility in completing multiple NCA tasks and flexibility in adapting to diverse scenarios.
We observe that key NCA tasks including configuration generation, translation, and analysis fundamentally involve text-to-text transformation. 
Both configuration generation and analysis transform text between device configuration and network intent, while configuration translation converts text from one vendor syntax to another. 
Such text-to-text essence creates opportunities for NCA tools with enhanced versatility.
Moreover, recent advances in natural language processing (NLP) have led to powerful language models (LM) for text comprehension and generation.
These models do not require extensive manual effort and domain expertise to adapt to new scenarios. 
Instead, they automatically extract linguistic patterns and features from data. 
This allows LMs to flexibly extend to new network protocols, devices, and applications without complex re-engineering.
Overall, by framing NCA tasks as text-to-text problems, we can apply LMs to unify and automate multiple functions including generation, translation, and analysis. 
This provides a versatile and unified approach for diverse NCA needs, in contrast to using multiple specialized tools. 
Our vision of utilizing LMs for NCA faces two key challenges in practice.
LMs are known to lack specialized domain knowledge. 
Network configuration is no exception. 
Existing research indicates that current LMs lack professional expertise for robust network configuration tasks~\cite{alan}, failing to achieve expected performance in this domain. 
Unfortunately, there is a scarcity of high-quality corpus data and supervision data in the network configuration field.
The main challenges in addressing this revolve around two aspects:
\begin{itemize}
\item LMs are not sensitive or accurate with respect to technical context and domain-specific terminology. 
The first challenge is thus how to obtain network configuration corpora and effectively integrate this professional knowledge into the model.
\item There is almost no task supervision data available in the network configuration domain. It is necessary to explore how to obtain or generate such task supervision data, and enable models to learn efficiently from diverse data formats to master multiple NCA tasks.
\end{itemize}
Overcoming the challenges is critical to unlocking the potential of LMs for flexible and versatile NCA. Our work aims to make progress on these fronts.


      



To address the first challenge of integrating specialized knowledge, we construct a network configuration corpus by extracting data from vendor manuals and community forums. 
We then continue pretraining a programming language model on this corpus via transfer learning. 
This enables efficient adaptation of the LMs to the specialized terminology and semantics of the network configuration domain.
To address the scarcity of supervision data for network configuration tasks, we present a novel data mining technique. Specifically, we leverage the robust language capabilities of LMs to intelligently mine configuration task data. The model interacts with a simulated environment to generate putative training examples, which are then validated against network analysis tools to ensure correctness. This provides an efficient way to accumulate the diverse supervision data needed for real-world configuration tasks.
With this expanded dataset, we train the model on these tasks in a multi-task learning framework. 
This equips the model with skills for generating, analyzing, and translating configurations across diverse networking environments.

Our key contributions are summarized as follows:
\begin{itemize}
\item This work is the first to model network configuration tasks, including generation, translation, and analysis, as text-to-text problems that can be tackled by LMs. We provide an analysis of the potential and challenges of this approach for automating network configuration.
\item We propose PreConfig, a novel pretrained language model tailored specifically for diverse NCA tasks through continued pretraining on in-domain corpora and fine-tuning on downstream tasks. This enables PreConfig to acquire abilities for robust configuration generation, multi-vendor translation, and semantic analysis.
\item Extensive experiments on representative NCA tasks demonstrate PreConfig's strong performance compared to conventional LMs. Quantitative and qualitative results show PreConfig produces network configurations that are more syntactically and semantically correct, while better capturing the nuances of complex networking environments.
\end{itemize}

The remainder of this paper is organized as follows. Section 2 introduces the background for this work. Intuition is presented in Section 3. Section 4 details the proposed solution of PreConfig. Section 5 presents experiments evaluating the effectiveness of PreConfig in configuration tasks. Related work is introduced in Section 6. Finally, we conclude this paper in section 7.


\section{Background}

In this section, we introduce tasks in the context of network configuration automation, followed by an introduction of robotic process automation with LMs.

\subsection{Network Configuration Automation}
\label{sec:2.1}

In recent years, with the continuous expansion of network scale and the enhancement of functional requirements, network operators have begun to realize the limitations of traditional network configuration methods in meeting the requirements of complex network operations. 
To simplify network configuration management, introducing automated methods has become increasingly critical~\cite{knowledge,ai}. 

\begin{figure}[tp]
    \centering
    \includegraphics[width=1\linewidth]{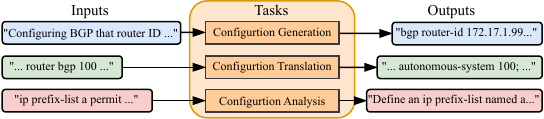}
    \caption{An example of text transformation for configuration tasks. Configuration generation and analysis tasks involve the transformation between network configuration and natural language, while the configuration translation task involves the transformation of configurations from different vendors.}
    \label{fig:fig1}
\end{figure}

\begin{figure}[tp]
    \centering
    \includegraphics[width=1\linewidth]{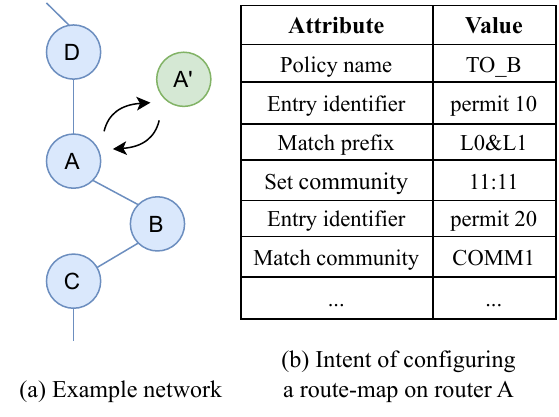}
    \caption{(a) is an example network, where A' represents a Juniper router used as a backup for Cisco router A. (b) is an example configuration intent for network operators, used to configure route-maps for router A.}
    \label{fig:fig0}
\end{figure}

\begin{figure*}[tp]
    \centering
    \includegraphics[width=0.8\linewidth]{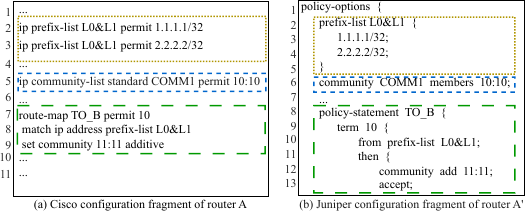}
    \caption{Cisco and Juniper configuration fragment of route maps. The parts highlighted in the same color in (a) and (b) represent configuration modules with the same functionalities from different vendors.}
    \label{fig:fig2}
\end{figure*}

In this context, the paper focuses on three key tasks in network configuration automation: configuration generation, translation, and analysis. 
We use the examples in Figures~\ref{fig:fig1}, ~\ref{fig:fig0} and~\ref{fig:fig2} to illustrate these concepts. 
Figure~\ref{fig:fig1} shows an example of text transformation for each task. 
Figure~\ref{fig:fig0}(a) shows an example network with four routers (A-D) and three links. 
Currently, to modify the route propagation, it is necessary to add configuration elements such as route policies and access control lists to router A. 
To avoid potential issues like route interruptions, or to meet requirements such as device replacement or preventing vendor monopolies, it is necessary to use a router A' from another vendor as a backup for router A.
For business requirements, it is essential to analyze the current configuration for enhanced operational efficiency and business adaptability. 
These three processes correspond to three network configuration tasks.
We elaborate on each of these tasks below.

{\bf Configuration generation.} Figure~\ref{fig:fig0}(b) illustrates an example of the configuration intent for router A. 
To control route selection and traffic forwarding, network operators need to create a routing policy. 
The table in Figure~\ref{fig:fig0}(b) describes the creation of a route-map, involving multiple statements to match prefix lists and perform corresponding actions. 
In the practical process of configuring network devices, network operators formulate configuration intent, as depicted in Figure~\ref{fig:fig0}(b). 
Following this, they input configuration commands relying on manuals or personal experience. 
Finally, the configuration files are generated by network devices, as illustrated in Figure~\ref{fig:fig2}(a). 
This process is undoubtedly laborious, particularly in the configuration of large networks.
For the configuration generation task, as shown in the blue section of Figure~\ref{fig:fig1}, our target is to achieve the transformation from natural language to network configuration.

{\bf Configuration translation.} As shown in the green section of Figure~\ref{fig:fig1}, the configuration translation task involves the transformation between configurations from different vendors. For the backup between Cisco router A and Juniper router A' in Figure~\ref{fig:fig0}(a), network operators need to translate the configuration from Cisco to Juniper and make sure that the functionality of the configuration remains consistent before and after translation.
We provide configuration texts for routers A and A' in Figure~\ref{fig:fig2}.
Both configuration paragraphs specify the prefix lists (lines 2-3 in Figure~\ref{fig:fig2}(a) and 2-5 in Figure~\ref{fig:fig2}(b)), community list (lines 5 in Figure~\ref{fig:fig2}(a) and 6 in Figure~\ref{fig:fig2}(b)) and routing policy (lines 7-9 in Figure~\ref{fig:fig2}(a) and 8-13 in Figure~\ref{fig:fig2}(b)).
It is evident that there are significant syntactic differences between configurations from different vendors.
This is a challenging task as the operational commands and logic between configurations of different vendors are not entirely consistent~\cite{campion}.

{\bf Configuration analysis.} Configuration analysis aims at assisting network operators in analyzing and extracting the functional description of device configuration. 
Network configuration can be complex and extensive. 
In the core routers of large networks, configuration files can reach thousands of lines. 
Unlike the simple routing policies depicted in Figure~\ref{fig:fig2}, the elements in real network configurations, such as routing policies and access control lists, interact with each other. 
This mutual interaction poses significant challenges for network operators in analyzing network behaviors. 
Configuration analysis task provides a comprehensive analysis of network configuration to extract configuration intent. 
As shown in the red section of Figure~\ref{fig:fig1}, different from formatted configuration information mining~\cite{batfish}, our target is to accomplish the transformation from network configuration to natural language through configuration analysis task.

\subsection{Robotic Process Automation with LMs}

Robotic Process Automation (RPA) is a form of process automation technology, designed to leverage machines, scripts, or other means to simulate human interactions for the purpose of automating processes.
Automating software engineering, automated management of operational logs, and network configuration automation can all be classified as RPA technologies.

LMs are a type of technology utilizing statistics and machine learning.
In recent years, with the development of technology in LMs, they have demonstrated excellent capabilities in text generation and semantic understanding.
This robust text processing capability present opportunities for text transformation tasks of various domains.

Existing research, such as ProAgent, PreSQL, and CodeT5, has incorporated LMs into RPA, making advancements in automation processes within their respective domains.

\section{Intuition}

Our core insight is that the NCA tasks described in section~\ref{sec:2.1} fundamentally involve text-to-text transformation. 
Such text-to-text essence empowers LMs to serve as potent tools for handling NCA tasks.  
We begin by exploring the performance of existing LMs in the configuration tasks. 
We observe that even the most advanced LMs struggle to fully comprehend complex network configuration knowledge. 
We then identify two key challenges in applying LMs to the configuration tasks.

\begin{table}[tp]
  \begin{center}
    \caption{Evaluation of GPT's capabilities in configuration generation and understanding tasks.}
    \label{tab:table1}
    \begin{tabular}{|c|c|c|c|} 
      \hline
      \textbf{Task} & \textbf{Model} & \textbf{Dataset size} & \textbf{Score} \\
      \hline
      Understanding & GPT-3.5 & 326 & 36.50\% (EM)\\
      \hline
      Generation & GPT-3.5 & 500 &    54.33\% (BLEU)\\
      \hline
      Understanding & GPT-4 & 326 & 37.73\% (EM)\\
      \hline
      Generation & GPT-4 & 500 &    57.82\% (BLEU)\\
      \hline
    \end{tabular}
  \end{center}
\end{table}

\subsection{Exploring the Capabilities of LMs to Configuration Tasks}
\label{sec:3.1}

Current state-of-the-art LMs are pretrained on text from a variety of sources. 
Consequently, LMs such as ChatGPT can generate common configuration commands. 
However, recent research indicates that when applied to tasks such as configuration translation and configuration synthesis, network configuration generated by GPT-4 exhibit syntax and semantic errors. 
Even with the inclusion of multi-turn human feedback, GPT-4 still fails to rectify all errors~\cite{gpt4}. 
To explore the capabilities of LMs in the network configuration field, we conduct an analysis through the following two aspects:
\begin{itemize}
\item The comprehension capability of LMs for configuration commands.
\item The ability of LMs in generating configuration.
\end{itemize}

To assess the configuration understanding capability of LMs, we collected 326 functional description texts of configuration commands related to Cisco's BGP and OSPF configuration from Cisco's online tutorials.
We input the functional description texts in the form of questions to GPT. 
If the generated configuration commands differ from the answer, we consider GPT as a lack of understanding of that command.
We employ the Exact Match (EM) to evaluate GPT's understanding capabilities of configuration commands, which involves calculating the proportion of correct configuration commands in GPT's responses over the test dataset. We present an example of a wrong response from GPT in Appendix~\ref{appendix:a}. 

To assess the configuration generation capability of LMs, we collected 500 configuration snippets along with their natural language descriptions from Cisco's online tutorials.
We input the natural language text in the form of questions to GPT and use BLEU, a machine translation assessment metric, to evaluate the similarity between the configuration snippets generated by GPT and the reference configuration.

\begin{figure*}[tp]
    \centering    \includegraphics[width=1\linewidth]{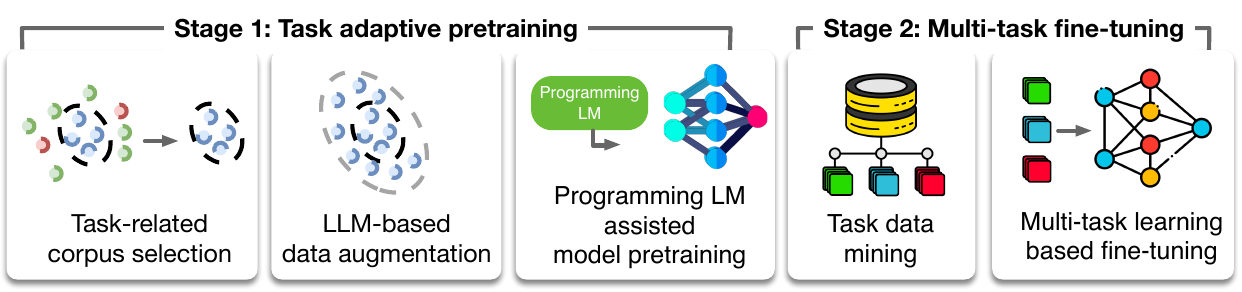}
    \caption{Overall framework of PreConfig. It consists of two stages. In stage 1, the pretraining data is collected through task-related corpus selection and LLM-based data augmentation. The pretrained model is obtained through continuous pretraining on a programming language model. In Stage 2, task data mining is accomplished through an LLM-based agent. 
The model acquires the capability to handle multiple tasks through multi-task learning based fine-tuning.}
    \label{fig:fig3}
  \end{figure*}

Table~\ref{tab:table1} displays the results, showing that GPT only achieved 37.73\% and 57.82\% at most on the evaluation metrics for the two tasks. 
We observe instances where GPT made errors in the order of keywords within configuration statements and confused configuration commands with similar concepts. 
This indicates that current LMs significantly lack comprehension and generation capabilities of network configurations.

\subsection{Challenges of Adapting LMs to NCA}
\label{sec:3.2}

Current LMs acquire task-specific capabilities by pretraining on task-relevant corpora and fine-tuning on high-quality task data. 
This insight motivates our research to explore the application of LMs to network configuration tasks. 
However, to achieve this target, we face two challenges.

\textit{How to obtain network configuration corpora and effectively integrate this professional knowledge into the model?}
Current LMs demonstrate robust comprehension of code and natural language after pretrained on open-domain corpora.
However, when it comes to specific domains like network configuration, LMs lack the capability to comprehend domain-specific languages. 
In this context, the first challenge we need to address is how to obtain network configuration corpora and effectively integrate them into the model. 
Manual data collection from vendor manuals or community forums is undoubtedly inefficient.
Therefore, we need to explore efficient methods for collecting data related to network protocols, configuration statements, and integrating them into the model. 
Moreover, due to security requirements, data is scarce in the field of network configuration. 
How to enable the model to comprehend configuration knowledge with scarce data is a crucial challenge to address.

\textit{How to obtain or generate task supervision data, and enable models to master multiple NCA tasks?} 
Learning on high-quality task supervision data is essential for LMs to handle specific tasks~\cite{biglog}. 
Different tasks have specific requirements for data.
For example, configuration generation and analysis tasks require transforming text of natural language and configuration language.
Configuration translation task, on the other hand, impose more stringent demands on data, requiring transforming configuration of different vendors. 
However, data mining for configuration tasks is difficult, especially for configuration translation task, where supervision data is almost nonexistent~\cite{nassim}. 
In addition, after obtaining task supervision data, we aim to develop a model capable of handling various tasks. 
How to enable the model to efficiently learn from data in various formats and master multiple tasks is a key challenge that we need to consider.

\section{The Proposed Solution}

Motivated by the limitations of existing NCA tools, we introduce PreConfig, a pretrained model to complete NCA tasks involving configuration generation, translation, and analysis. 
The overall framework of PreConfig is shown in Figure~\ref{fig:fig3}.
The implementation of PreConfig consists of the following two stages:
\begin{itemize}
\item Configuration corpus construction  and model pretraining.
\item Task data mining and model fine-tuning.
\end{itemize}

\subsection{Overview}
In the first stage, we build an automated framework for collecting configuration knowledge and pretraining the model.
The text of vendor manuals and community forums contains a wealth of knowledge in the field of network. 
However, the target of our knowledge collection is to extract configuration snippets and natural language text that are more relevant to configuration tasks from the extensive corpora of vendor manuals and community forums.
Because task-adaptive pretraining, as opposed to domain-adaptive pretraining, with a smaller-scale dataset, can more directly enhance the model's performance on downstream tasks~\cite{dont}.
After in-depth research into manuals and online tutorials provided by major vendors, we summarize paragraphs and webpage tags containing configuration corpus.
After that, we develop specific parsers to collect configuration corpus from web pages.
However, constructing parsers is challenging for vendor manuals and community forums with diverse data formats. Configuration snippets often coexist with natural language text, making it difficult to separate through manual and algorithmic design.
To address this challenge, we abstract the problem into a text classification task, utilizing a bag-of-words language
model to separate configuration snippets from mixed-language corpora.
Moreover, in view of the powerful text processing capabilities of LMs, we explore utilizing them to expand the collected configuration data.
Finally, due to the scarcity of configuration data, we employ the method of transfer learning. 
We inject the configuration corpus into a programming language model through continued pretraining to enhance the model's comprehension of configuration knowledge.

In the second stage, we build an LLM-based agent to obtain task data and utilize a multi-task learning framework to empower the model with the ability to handle configuration tasks.
For the acquirsion of task supervision data, traditional methods heavily rely on manual efforts.
In the field of software engineering, despite the existence of open-source programming language data, code experts still spend a lot of effort on mining data for tasks such as code generation, analysis, and translation~\cite{domain}.
For the configuration domain with scarce data, relying on manual data mining is evidently a challenge for us.
To address this challenge, we design an intelligent agent to simulate manual data mining.
Our approach involves utilizing LMs to interact with a simulated environment to generate putative task data and validating the data using configuration analysis tools to ensure its correctness.
In addition, research in NLP has demonstrated that multi-task learning can improve the performance and generalization of the model across tasks. 
Inspired by this idea, we train the model on generation, translation, and analysis tasks in a multi-task framework, empowering the model to handle diverse tasks.
In the next three sections, we provide a detailed implementation of PreConfig.

\begin{algorithm}[tp]
    \SetAlgoNoLine
    \caption{Task-related Corpus Selection Algorithm.}
    \label{algorithm}
    \KwData{User community website $L$, high quality configuration data $D2$}
    \KwResult{Generated datasets $D$ (a list of saved data)}
    $D \leftarrow []$\;
    $D1 \leftarrow \text{DataProcess}(L)$\;
    $model \leftarrow \text{ModelPretrain}(D1, D2)$\;
    \For{$d \in D2$}{
        $candidates \leftarrow \text{DataSelection}(d, model, n)$\;
        \For{$c \in candidates$}{
            $D \leftarrow D.\text{append}(\text{DataJudgment}(c))$\;
        }
    }
\end{algorithm}

\subsection{Configuration Corpus Construction and Model Pretraining}
\label{sec:5}
In this section, we present our method of constructing configuration corpus and the pretraining of PreConfig.

\subsubsection{Task-related corpus selection} 
\label{sec:5.1}
The target of configuration corpus construction is to collect configuration snippets related to configuration tasks.
For online tutorials on webpages, we extract configuration snippets by designing HTML parsers.
For vendor manuals and community forums, we design a task-related corpus selection algorithm based on a bag-of-words language model~\cite{vampire}.
We describe the workflow in Algorithm~\ref{algorithm}.
The algorithm mainly consists of three steps:
\begin{itemize}
\item The DataProcess function extracts user discussion text from the URL L. This function further preliminarily separates natural language text from configuration snippets and returns the initial community data D1.
\item The ModelPretrain function pretrains a bag-of-words model using data D1 and standard configuration snippets D2 to obtain embeddings for all data.
\item For each standard configuration snippet d in D2, the DataSelection function selects the top n most similar candidate texts from the overall data utilizing a K-Nearest Neighbors (KNN) algorithm. The DataJudgment function then returns the filtered configuration snippets.
\end{itemize}
Through the above method, we efficiently collect data and build a configuration corpus. 
We also apply the same approach to collect and process natural language text related to configuration tasks.

\subsubsection{LLM-based Data Augmentation}
\label{sec:5.2}
Leveraging LMs for text process can be a novel and practical method for data augmentation. 
It can effectively expand the scale of data while improving data diversity~\cite{databalance,databalance1}.
After completing configuration data collection, we propose a novel method for configuration data augmentation utilizing LMs.
How to ensure the quality of configuration text generated by the LMs is a core issue for us. 
We address this challenge by designing various detailed prompt templates.

\begin{figure}[tp]
    \centering    
    \includegraphics[width=1\linewidth]{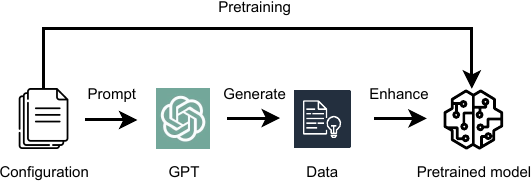}
    \caption{Implementation of configuration data augmentation utilizing prompt engineering. The inputs to GPT are configuration snippets, prompt templates, and the outputs of GPT are expanded configuration data.}
    \label{fig:fig4}
  \end{figure}

As shown in Figure~\ref{fig:fig4}, we utilize prompt engineering to guide GPT to expand configuration data.
Firstly, we design prompt templates to present detailed task requirements to GPT. 
As shown in Table~\ref{tab:table2}, we incorporate domain knowledge into the prompt templates~\cite{prompt}, including vendor names, configuration attributes, and model roles.
Secondly, we employ Standard Operating Procedures (SOP) to standardize the operation process of the GPT, generating high-quality configuration data~\cite{metagpt}.
Finally, we use the data generated by GPT and the original data together as the pretraining data for the model.

\begin{table}[tp]
  \begin{center}
    \caption{The design of prompt templates for configuration data augmentation.}
    \label{tab:table2}
        \begin{tabular}{|l|l|} 
          \hline
          \textbf{Methods} & \textbf{Prompt Templates}\\
          \hline
          \multirow{3}{*}{Raw} &  Please help me enhance this configur-\\
          & ation text:\\
          & \{INPUT\_CONFIG\}\\ 
          \hline
          \multirow{4}{*}{Raw+DSP} &  You are an expert in network configu-\\
          & ration domain, Please help me enhanc-\\
          & e this configuration text:\\
          & \{INPUT\_CONFIG\}\\
          \hline
          \multirow{6}{*}{Raw+DSP+SOP} &  You are an expert in configuration d-\\        
          & omain, Please help me enhance this c-\\
          & onfiguration text, considering vario-\\
          & us combinations of protocol paramete-\\
          & rs and statements:\\
          & \{INPUT\_CONFIG\}\\
          \hline
        \end{tabular}
  \end{center}
\end{table}

\begin{table*}[tp]
  \begin{center}
    \caption{Example noisy inputs and Original text sequence during pretraining of PreConfig.}
    \label{tab:table3}
        \begin{tabular}{|l|l|} 
          \hline
          \textbf{Noisy Input} & \textbf{Original text Sequence}\\
          \hline
          BGP uses a [MASK] ID to identify BGP-speaking [MASK] . & <nl> BGP uses a router ID to identify BGP-speaking peers.\\
           <nl> & \\
          \hline
          bgp \{ group \{ type ; import Default ; export ; peer-as 100 ; & <juniper> bgp \{ group \{ ISP-AS100 \{ type external ; import\\
          neighbor \{ description " ISP FastAccess: Circuit GD8AJ12B: & Default ; export Direct-To-BGP ; peer-as 100 ; neighbor \\
          ISP NOC 800-111-2222 " ; \} \}... <juniper> & 120.0.4.9 \{ description " ISP FastAccess: Circuit GD8AJ12B:\\
            & ISP NOC 800-111-2222 " ; \} \}...\\
          \hline
          router ospf 104 NEW\_LINE redistribute bgp 104 subnets & <cisco> router ospf 104 NEW\_LINE redistribute bgp 104\\
          NEW\_LINE network 104.0.0.0 0.0.0.255 [MASK] <cisco> & subnets NEW\_LINE network 104.0.0.0 0.0.0.255 area 0\\
          \hline
        \end{tabular}
  \end{center}
\end{table*}

\subsubsection{Pretraining of PreConfig}
In this section, we describe the details of pretraining, including pretraining data, input-output representations, model architecture, and pretraining tasks.

{\bf Pretraining data.} We use the methods introduced in~\ref{sec:5.1} and~\ref{sec:5.2} to collect data for pretraining PreConfig, including Cisco and Juniper configuration snippets, as well as natural language text related to configuration tasks. 
The statistics of the pretraining data are presented in~\ref{sec:8.1}.

{\bf Input-output representations.} PreConfig utilizes text in three languages for pretraining: configuration language (Cisco and Juniper), and natural language (English). As shown in Table~\ref{tab:table3}, for each language, we add a language tag (e.g., Cisco or Juniper) in its input sequence to distinguish between different languages.

{\bf Model architecture.} PreConfig uses the transformer architecture, the same as BART. 
Leveraging the thought of transfer learning, we initialize PreConfig with pretrained parameters of PLBART, a programming LM, to accelerate model pretraining and enhance its comprehension of configuration languages.

{\bf Model pretraining.} PreConfig is pretrained by corrupting text and optimizing reconstruction loss.
As shown in Table~\ref{tab:table3}, we use three strategies including token masking, token deletion, and token infilling to pretrain PreConfig. 
Token masking randomly samples input tokens and replaces them with [Mask]. 
Token deletion randomly removes input tokens. 
Token infilling replaces a consecutive sequence of tokens with a single [Mask].

The reconstruction loss of PreConfig, that is, the cross entropy between the output of decoder and the ground truth, is as follows:
{\setlength\abovedisplayskip{0.1cm}
\setlength\belowdisplayskip{0.1cm}
\begin{equation}
\mathcal{L}(\theta) = \sum_{i=1}^{n} \log{P(X_i|f(X_i);\theta)}
\end{equation}}
Where $\theta$ is initialized with the PLBART parameters, \emph{n} is a set of pretraining data, and \emph{f} is a function that generates noisy text.

\subsection{Task Data Mining and Model Fine-tuning}
\label{sec:6}
In this section, we design an intelligent agent, mimicking the human data mining process to complete automated task supervision data mining.
After that, we train the model utilizing a multi-task learning framework to enable it to handle multiple configuration tasks.

\subsubsection{Task Data Mining}
\label{sec:6.1}
\begin{figure}
    \centering
    \includegraphics[width=1\linewidth]{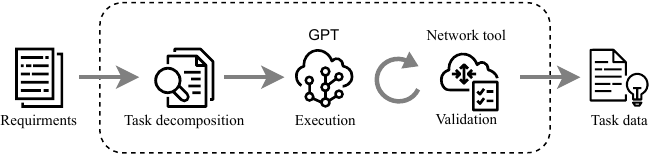}
    \caption{ConfigExtract: An intelligent agent for multi-task supervision data mining. The input is the data mining requirements for a specific task, the output is task supervision data.}
    \label{fig:fig5}
\end{figure}

Recently, research on intelligent agents has received widespread attention, aiming to emulate human work in real-world scenarios.
Inspired by this idea, referring to methods such as Chain-of-Thought (CoT)~\cite{cot} and behavior simulation~\cite{behave}, we design ConfigExtract, a tool that mimics domain experts for mining configuration tasks supervision data.
As shown in Figure~\ref{fig:fig5}, ConfigExtract first decomposes a complex task into multiple sub-tasks that are easy to execute and manage. 
Then, it completes the sub-tasks and returns preliminary results. 
Finally, it ensures data correctness through interaction with network analysis tools and human feedback.

{\bf Task decomposition.}
The process from requirements input to the data output involves numerous subtasks, involving multiple text processing steps.
In this paper, we focus on data mining for configuration generation, translation, and analysis tasks. 
The distinct requirements determine different text processing steps for each task.
After determining the task requirements, ConfigExtract decomposes them into multiple text processing steps. 
In this process, we incorporate detailed domain information constraints, specifying the input-output for each step and the requirements for text transformation, to ensure the correctness of intermediate steps and the results.

{\bf Execution and supervision.}
After completing the task decomposition, ConfigExtract processes the input text, accomplishes all subtasks, and returns results. 
To prevent it from entering into an infinite loop, we establish a supervision mechanism. 
ConfigExtract conducts self-checks on the output. 
For incorrect outputs, ConfigExtract marks them and returns them for manual correction.

{\bf Validation and feedback.}
To ensure the correctness of the data generated by ConfigExtract, we utilize configuration analysis tools for verification. 
To ensure the grammatical correctness of configuration data, we utilize Batfish to check the syntax of the generated configuration. 
Additionally, for configuration translation task, we utilize Campion to ensure semantic consistency across configurations from different vendors.
If the generated data is incorrect, ConfigExtract will provide feedback to GPT and iteratively modify the data until it is error-free.

\subsubsection{Fine-tuning of PreConfig} 
After obtaining task supervision data, we leverage a multi-task learning framework to train the model to handle multiple configuration tasks. 
Multi-task learning (MTL) is a machine learning method that aims to enhance the performance and generalization ability of a model by training it to accomplish multiple related tasks.
In traditional single-task learning, models focus on a specific task. 
However, in the real world, many tasks share correlated features, and multi-task learning can leverage this correlation to enhance the model's performance.
In the fields of NLP and programming language processing, research like T5 and CodeT5~\cite{t5,codet5} has proved that multi-task learning can improve a model's performance on downstream tasks.

To train the model for multiple configuration tasks, we unify the data formats and add language tags at the beginning of the data.
To mitigate the impact of imbalanced dataset sizes on different tasks, we employ a balanced sampling algorithm.
Specifically, for \emph{N} task datasets with probabilities \{\emph{q$_1$, q$_2$,...,q$_N$}\}, we sample by the distribution below:
{\setlength\abovedisplayskip{0.1cm}
\setlength\belowdisplayskip{0.1cm}
\begin{equation}
q_i=\frac{r_i^\alpha}{\sum_{j=1}^N r_j^\alpha},r_i=\frac{n_i}{\sum_{k=1}^N n_k}
\end{equation}}
where \emph{$n_i$} is the number of examples of the i-th task, and $\alpha$ is set to 0.5.

\subsection{Intent Representation and Generalization}
\label{sec:7}

Configuration generation and analysis tasks require the transformation from intent to configuration text~\cite{intentdriven}.
The configuration intent of PreConfig is to represent network operators' descriptions of protocol attributes and parameters for device configuration.
Configuration is fixed and precise, while intent may vary in expression due to factors such as the operator's personal experience.
This situation is also present in network configuration tools, such as config2spec and Aura~\cite{config2spec,aura}, where various policy languages have been designed to express the configuration intent.
The target of PreConfig is to complete the transformation from natural language text to single device configuration. 
How to unify the mapping from natural language to configuration is a key challenge.
In this section, we present our exploration and insight, introducing our method of representing intent for configuration and utilizing LM to generalize intent representations.

\subsubsection{Intent Representation}
In real-world scenarios, network operators record configuration intents using structured data.
As shown in Figure~\ref{fig:fig0}(b), network operators use tables to store attributes and parameters of routing policies.
Compared to this approach, our target is to express intents through natural language. 
This not only simplifies the configuration process but also provides a more intuitive means of interaction.
To achieve this target, we introduce our method of representing configuration intents.
Firstly, we extract key content, such as protocol attributes and parameters, from configuration using network analysis tools (e.g., Batfish). 
Subsequently, we design templates to transform this information into natural language text. 
Finally, we utilize LM to refine the natural language representation of configuration intent.

\subsubsection{Intent Generalization}
The natural language expression of network configuration intents is flexible and variable, incorporating specific experience of network operators. 
Additionally, the expression of configuration attributes varies among different vendors.
To enable PreConfig to accurately understand the configuration intent, we generalize the natural language text utilizing LMs.
Leveraging the robust understanding and generation capabilities of LMs for natural language, we utilize them to rephrase intent descriptions and present the output in an imperative statement form.
By generalizing natural language text, we empower PreConfig with a more robust understanding of configuration intent.

\section{Evaluation}
\label{sec:8}
In this section, we present the implementation of PreConfig, configuration tasks, datasets, experiment setup, and results.

{\bf Implementation}
PreConfig is implemented in Python.
We utilize the pretrained PLBART model with 406 million parameters as our base model.
The PLBART model is pretrained using extensive programming language and natural language data, resulting in powerful language comprehension and generation capabilities.
Due to the similarity between configuration language and programming language, the model can rapidly comprehend configuration language, contributing to its performance in configuration tasks. 
Through the application of transfer learning, we pretrain the PLBART model with configuration snippets and natural language text containing extensive configuration knowledge. 
Subsequently, we utilize a multi-task learning framework to fine-tune the model on configuration tasks.
In the end, we accomplish the implementation of PreConfig.
During the pretraining of PreConfig, we set the model's batch size to 2048, the learning rate to 3e-4, and utilize Adam for optimization.

\subsection{Configuration Tasks}
\label{sec:8.1}
{\bf Configuration generation} involves generating configuration for a specific vendor according to the natural language text (English). 
We set Cisco configuration as the output and evaluate the model's performance on this task utilizing evaluation metrics such as BLEU, ROUGE, and EM.

{\bf Configuration analysis} is the opposite of configuration generation, which involves transforming configuration into natural language text.
We take Cisco configuration as input and evaluate the quality of the generated natural language text.

{\bf Configuration translation} involves translating configuration into equivalent configuration for another vendor.
We evaluate the model's performance on translation task utilizing Cisco and Juniper configurations.

{\bf Evaluation metrics.} For task evaluation, we employ machine translation metrics such as BLEU, ROUGE, and EM to evaluate the model's performance in configuration tasks. 
BLEU measures the transformation based on precision, while ROUGE assesses transformation based on recall.
BLEU focuses on the accuracy of the output results, while ROUGE emphasizes the comprehensiveness of the output information.
EM measures the proportion of output that exactly matches the reference.

\subsection{Datasets}
\label{sec:8.2}
We obtain the data for model training through the methods described in~\ref{sec:5} and~\ref{sec:6}.
Specifically, we collected 7GB of vendor manuals and community forums data. 
Through the method in~\ref{sec:5}, we extracted 100MB of data for pretraining. 
Our pretraining data consists of Cisco and Juniper configuration snippets, along with task-related natural language text.
The final statistics for the pretraining and task datasets of the model are presented in Table~\ref{tab:table4}.
Our task datasets include configuration related to BGP, OSPF, static route, route policy, ACL, and other elements. 
We can also use the same methods to expand and support additional configuration elements.

\begin{table}[tp]
  \begin{center}
    \caption{Statistics of the pretraining datasets and tasks datasets.}
    \label{tab:table4}
    \begin{tabular}{|c|c|c|c|c|} 
      \hline
      \textbf{Task} & \textbf{Language} & \textbf{Train} & \textbf{Valid} & \textbf{Test}\\
      \hline
      \multirow{3}{*}{Pretrain} & cisco & 50000 & 6000 & 6000 \\
      & juniper & 42000 & 5000 & 5000 \\
      & nl & 54000 & 6500 & 6500 \\
      \hline
      Generation & nl to cisco & 4000 & 500 & 500\\
      \hline
      Analysis & cisco to nl & 4000 & 500 & 500\\
      \hline
      Translation & juniper to cisco & 2400 & 300 & 300\\
      \hline
    \end{tabular}
  \end{center}
\end{table}

\subsection{Experiment Setup and Results}
{\bf Experiment setup}
For the three configuration tasks, we explore the performance of the model after pretraining and fine-tuning in each task.
Subsequently, we investigate the impact of multi-task learning on the model's performance in configuration tasks.
Finally, we compare the performance of PreConfig with generic language models on tasks such as configuration generation.

\begin{figure}[tp]
    \centering
    \includegraphics[width=1\linewidth]{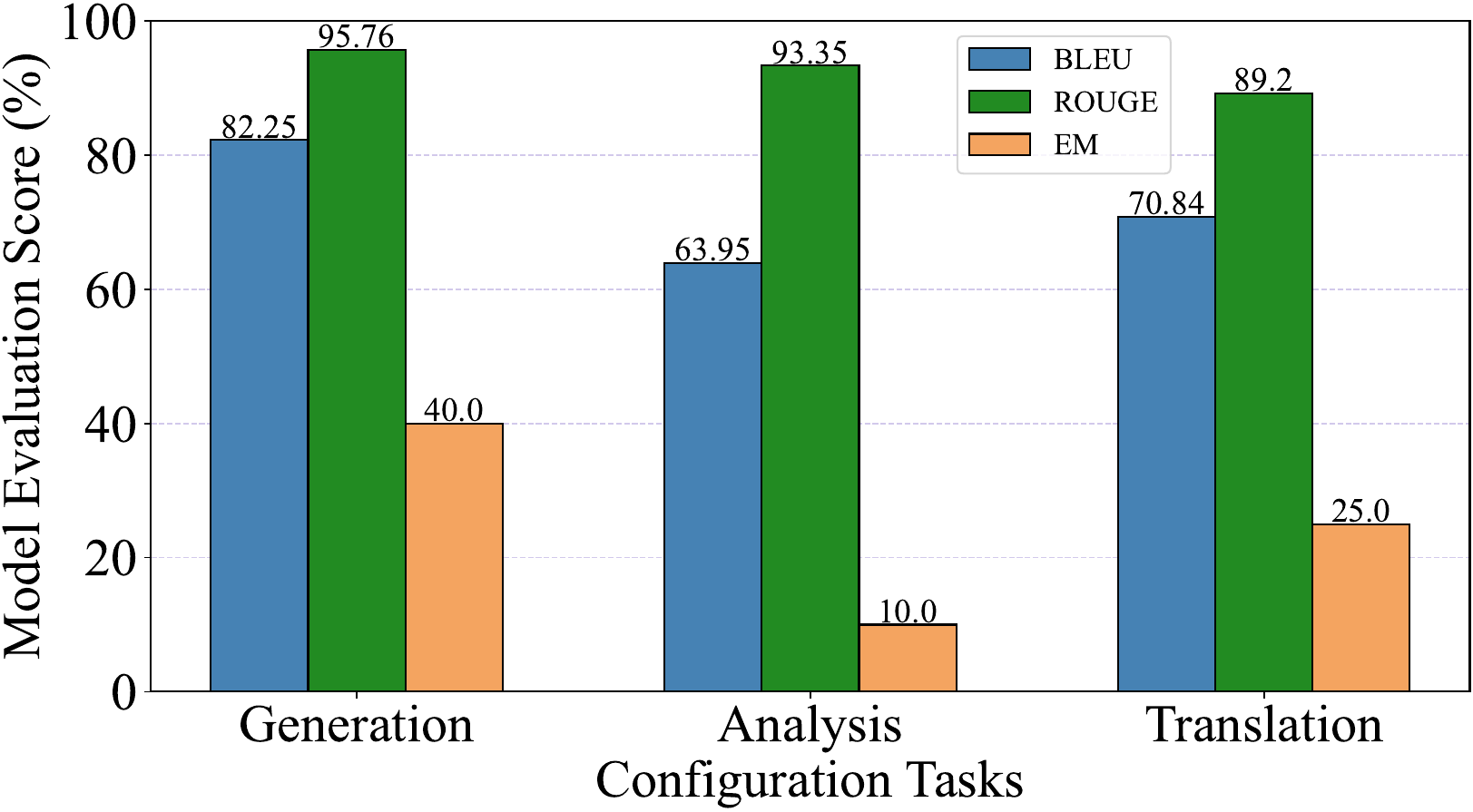}
    \caption{Evaluation results of single-task fine-tuning after pretraining. The evaluation metrics include BLEU, ROUGE, and EM.}
    \label{fig:fig6}
  \end{figure}

\begin{table}[tp]
  \begin{center}
    \caption{An example of configuration translation task. The input, reference, and the output of PreConfig are shown in three examples, respectively.}
    \label{tab:table5}
        \begin{tabular}{|l|l|} 
          \hline
          \textbf{Type} & \textbf{Configuraion}\\
          \hline
          \multirow{6}{*}{Input} &  routing-options~ \{\\
          & \ \ static~ \{\\
          & \ \ \ \ route~ 0.0.0.0/0~ next-hop~ 80.0.0.2;\\ 
          & \ \ \ \ route~ 0.0.0.0/0~ next-hop~ 80.0.0.1;\\
          & \ \ \}\\
          & \}\\
          \hline
          \multirow{2}{*}{Target} &  ip~ route~ 0.0.0.0~ 0.0.0.0~ 80.0.0.2\\
          & ip~ route~ 0.0.0.0~ 0.0.0.0~ 80.0.0.1\\
          \hline
          \multirow{2}{*}{PreConfig} &  ip~ route~ 0.0.0.0~ 0.0.0.0~ 80.0.0.2\\
          & ip~ route~ 0.0.0.0~ 0.0.0.0~ 80.0.0.1\\
          \hline
        \end{tabular}
  \end{center}
\end{table}

{\bf Downstream tasks.} We first explore the performance of the model in each configuration task after pretraining.
We continuously pretrain the PLBART model utilizing configuration snippets (Cisco and Juniper) and task-related natural language text (English).
Subsequently, we perform task-specific fine-tuning on the model for configuration generation, analysis, and translation. 
We utilize relevant evaluation metrics to assess the model's performance in each configuration task.
Figure~\ref{fig:fig6} illustrates the evaluation results of the model in each task.

For configuration generation task, the model achieves BLEU and ROUGE scores of 82.25 and 95.76, respectively, on the test dataset.
The model demonstrates a strong understanding of configuration language.
In addition, the model achieves an EM score of 40.0, indicating its capability to generate precise configuration.

For configuration analysis task, the model achieves BLEU and ROUGE scores of 63.95 and 93.35, respectively, while the EM metric is only 10.0.
This is because natural language is more flexible and varied in terms of grammar and semantic information compared to configuration language.
Therefore, even though the model is trained on the same data as the configuration generation task, the evaluation metrics for the analysis task significantly decrease.

For configuration translation task, the model achieves BLEU and ROUGE scores of 70.84 and 89.2, respectively, while the EM metric is 25.0.
Table~\ref{tab:table5} presents an example translation of static route.
It can be considered that after fine-tuning, the model has learned the mapping between configuration commands from different vendors, enabling it to obtain accurate translation results.

It is evident that, after high-quality network configuration knowledge injection and downstream task fine-tuning, the model has acquired the capability to handle each configuration task.

\begin{figure}[tp]
    \centering
    \includegraphics[width=1\linewidth]{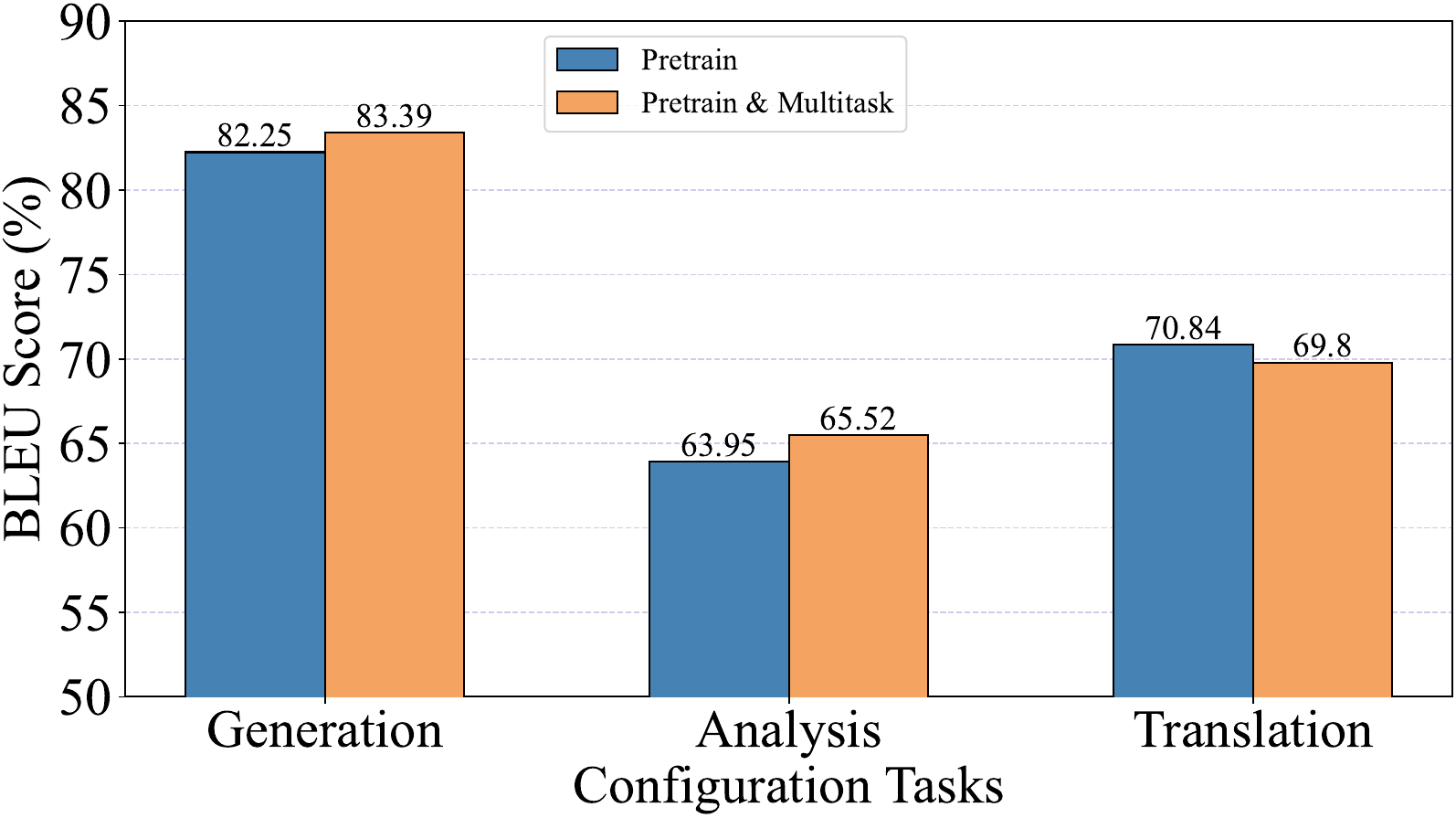}
    \caption{The impact of multi-task learning on the model's performance across three tasks. The BLEU metric is utilized for evaluation.}
    \label{fig:fig7}
  \end{figure}

\begin{figure}[tp]
    \centering
    \includegraphics[width=1\linewidth]{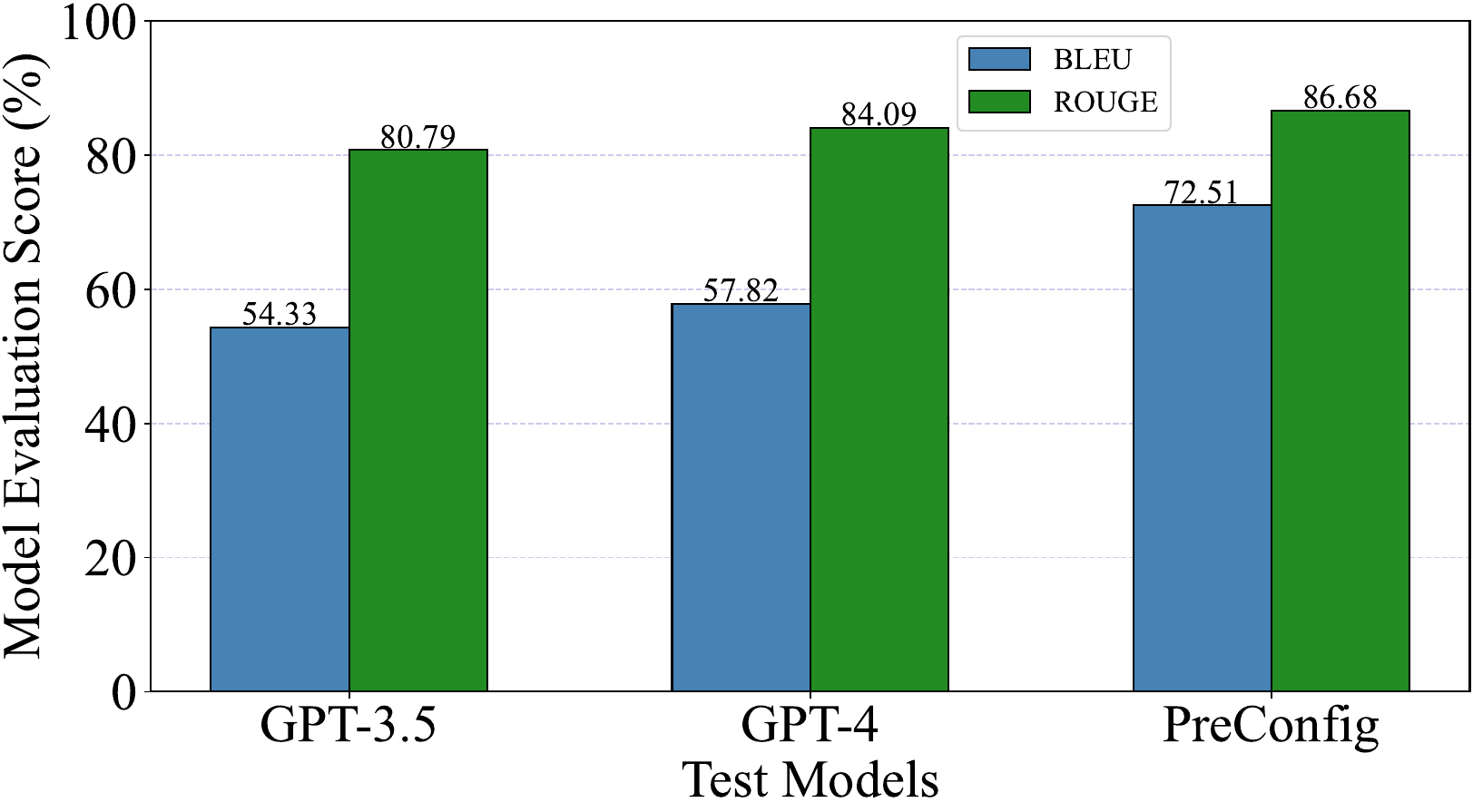}
    \caption{Comparison of evaluation results for PreConfig and GPT on configuration generation tasks. BLEU and ROUGE are used for evaluation.}
    \label{fig:fig8}
  \end{figure}

\begin{figure}[tp]
    \centering
    \includegraphics[width=1\linewidth]{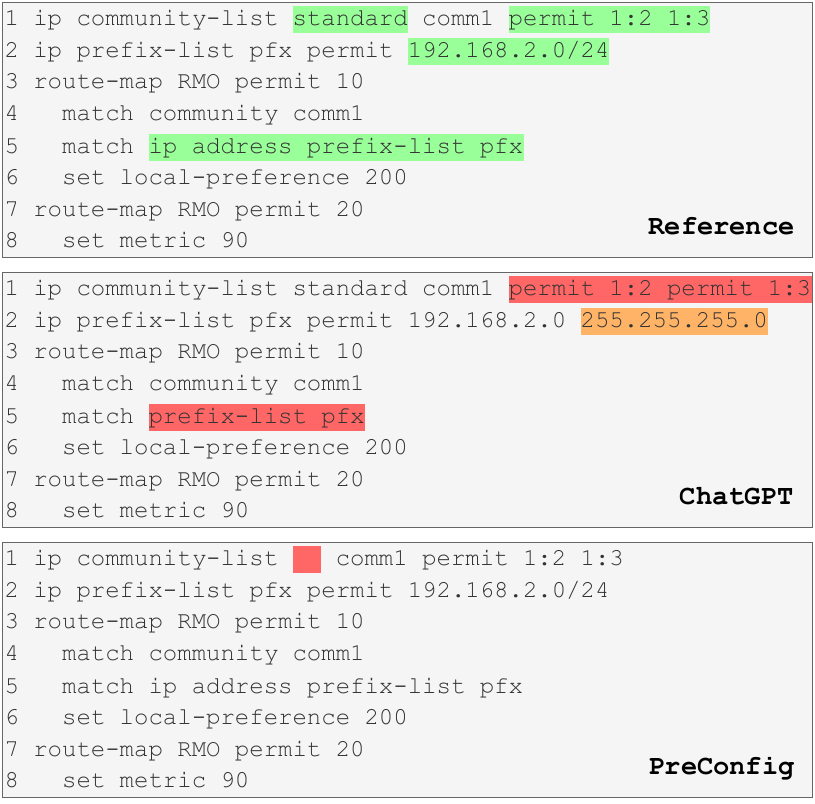}
    \caption{An example output of ChatGPT and PreConfig in configuration generation task. The content marked in red represents incorrect configuration, while the content marked in yellow indicates differing configurations.}
    \label{fig:fig9}
  \end{figure}

{\bf Effects of multi-task learning.}
After continued pretraining, we perform multi-task fine-tuning on the model. 
Multi-task learning involves fine-tuning a model on multiple tasks, leveraging the sharing of model parameters and training data to enhance the model's generalization and task-handling capabilities.
We conduct multi-task fine-tuning for configuration generation, translation, and analysis tasks.
The evaluation results for each task after multi-task fine-tuning are illustrated in Figure~\ref{fig:fig7}.
By comparing the results with single-task fine-tuning, we analyze the impact of multi-task learning on the performance of each task.
The model shows improvements in configuration generation and configuration analysis tasks, and it achieves performance in the configuration translation task similar to that of single-task fine-tuning.
Multi-task learning not only enables the model to handle multiple configuration tasks but also enhances the model's performance across different tasks.

{\bf Comparisons with generic language model.}
we compare the evaluation results of PreConfig with generic language models on tasks such as configuration generation.
We utilize the test dataset of configuration generation task collected in ~\ref{sec:3.1} as input for PreConfig. 
We calculate the BLEU and ROUGE scores of PreConfig and compare the results with those of GPT-3.5 and GPT-4.
As shown in Figure~\ref{fig:fig8}, compared to GPT-4, PreConfig demonstrates improvements of 14.69 in BLEU and 2.59 in ROUGE for the configuration generation task.
Figure~\ref{fig:fig9} shows an example of configuration generation task. 
In this case, we task both PreConfig and ChatGPT to accomplish the same assignment:
\begin{itemize}
\item Create a community-list named comm1, permit routes with community values 1:2 and 1:3.
\item Create an IP prefix list named pfx, permit routes matching 192.168.2.0/24. 
\item Create a route-map named RMO with sequence number 10, match community-list comm1 and prefix-list pfx, and set localpreference to 200. Create a route-map, named RMO with sequence number 20, and set metric to 90.
\end{itemize}
Figure~\ref{fig:fig9} shows that ChatGPT exhibits obvious syntax and semantic errors in configuring community-list and route-map. 
In contrast, PreConfig avoids such errors.
This indicates that compared to the generic language model, PreConfig can generate configuration with more accurate syntax and semantics. 
We also present a case of configuration translation task in Appendix~\ref{appendix:b}.

\section{Related Work}
\subsection{Language Models for Automating Software Engineering}
In recent years, with the development of machine learning, the field of software engineering has embarked on the integration of common tasks in program understanding and generation, such as code generation, code summarization, and code translation, with NLP techniques.
Programming language models have achieved excellent performance in these tasks.

CodeBERT~\cite{codebert} employs the same transformer-based encoder architecture as the BERT model. It undergoes pre-training using both programming language and natural language data and is utilized for tasks such as code search and code summarization.
CodeGPT~\cite{codegpt} utilizes the same transformer-based decoder architecture as the GPT model. 
It undergoes pretraining with programming language data and is employed for tasks such as code completion and code generation.
PLBART~\cite{plbart} employs a transformer-based encoder-decoder architecture similar to the BART model. It is used for common code generation and understanding tasks.

\subsection{Network Configuration Synthesis}
The target of network configuration synthesis is to automatically generate network configuration according to the user intent.
This involves transforming a set of high-level policies represented in a domain-specific language into low-level network configurations, thereby avoiding the occurrence of low-level errors during the manual configuration process.

NetComplete~\cite{netcomplete} is an incremental configuration synthesis tool that utilizes SMT solvers to resolve configuration parameters.
AED~\cite{aed} can perform incremental configuration synthesis and repair existing configurations, and AED can support some soft-constrained management objectives.

\section{Conclusion}

This paper introduces PreConfig, a pretrained  model for NCA tasks.  
First, PreConfig acquires expertise in network configuration through an automated framework for configuration corpora construction and model pretraining.
Second, we design an intelligent agent to mine configuration task supervision data. 
Through a multi-task learning framework, PreConfig acquires the capabilities to handle various configuration tasks.
Compared to current NCA tools, PreConfig can handle multiple configuration tasks and extend to complex application scenarios.
Extensive experiments demonstrate PreConfig's strong performance in configuration generation, translation, and analysis tasks. 
Compared to generic language models, PreConfig exhibits a more powerful capability in handling configuration tasks.


\noindent{\bf Ethical issues.} This work does not raise any ethical issues.

\balance
\bibliographystyle{ACM-Reference-Format}
\bibliography{reference}

\newpage
\appendix
\section{Example of Testing ChatGPT about Configuration Understanding}
\label{appendix:a}
\begin{figure}[tp]
    \centering
    \includegraphics[width=1\linewidth]{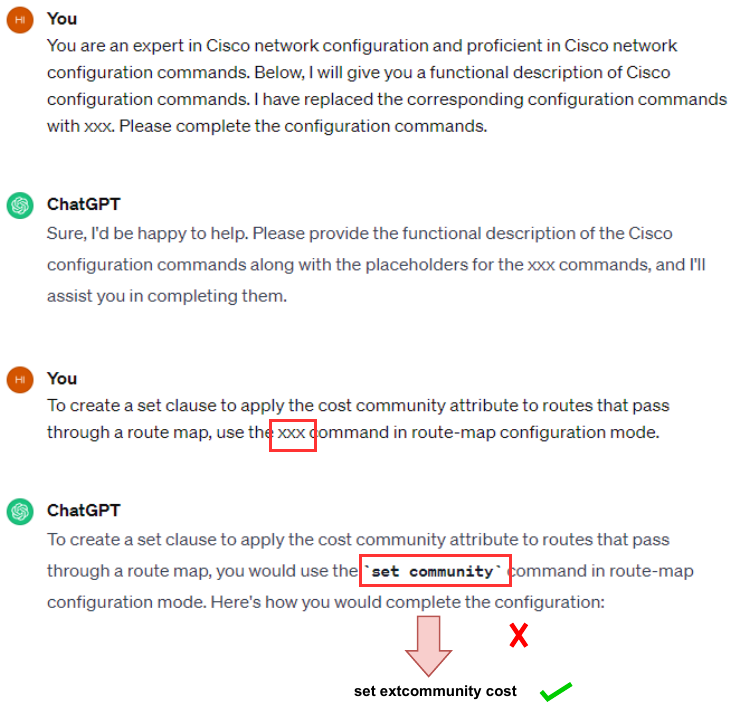}
    \caption{An example about the question of configuration understanding and incorrect answer from GPT.}
    \label{fig:fig10}
\end{figure}

Figure~\ref{fig:fig10} shows an example of a configuration understanding question we collected and the incorrect answer from GPT. 
The first line is our prompt, used to specify the task to be completed by GPT. 
The last line is the error response returned by GPT, indicating that GPT tends to confuse configuration commands with similar functionalities.

\section{Comparision of PreConfig with ChatGPT in Configuration Translation Task}
\label{appendix:b}
\begin{figure}[tp]
    \centering
    \includegraphics[width=1\linewidth]{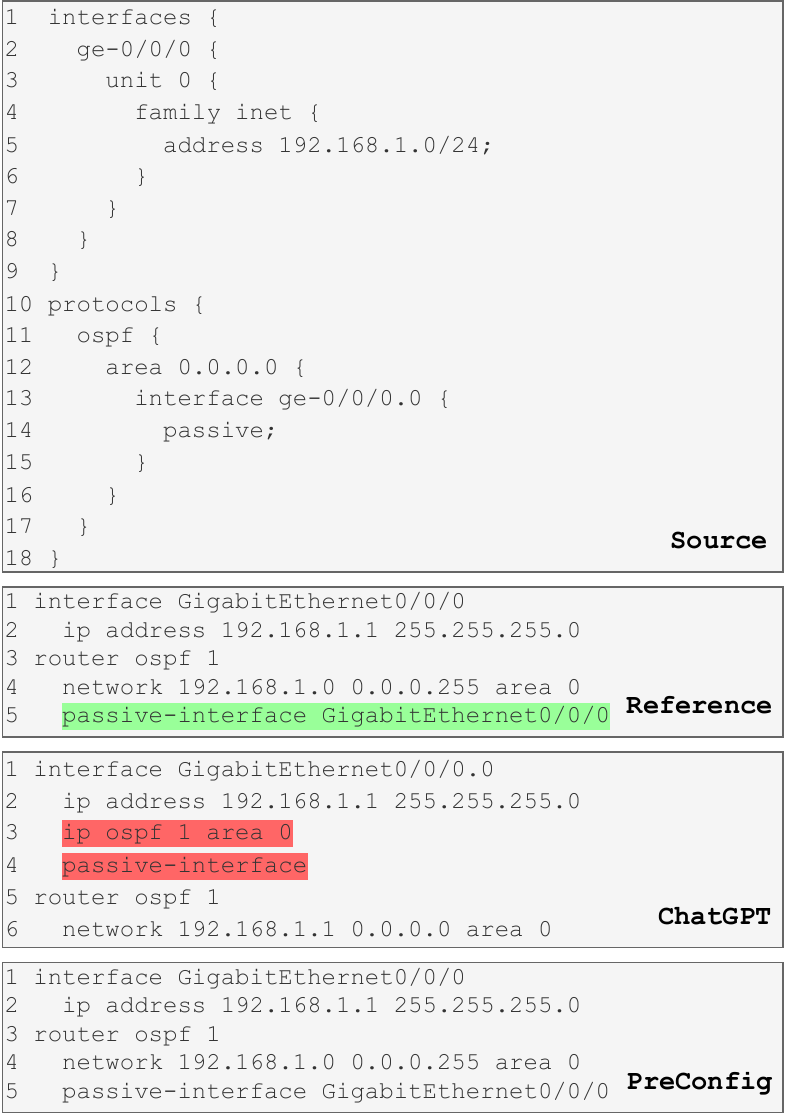}
    \caption{An example output of ChatGPT and PreConfig in configuration translation task. The content marked in red represents incorrect configuration, while the content marked in yellow indicates differing configurations.}
    \label{fig:fig11}
\end{figure}

Figure~\ref{fig:fig11} shows an example of the translation results of PreConfig and ChatGPT.
It shows errors in ChatGPT's configuration of passive interfaces.

\end{sloppypar}
\end{document}